\documentstyle[12pt]{article}
\begin{document}
\def\thebibliography#1{\section*{REFERENCES\markboth
{REFERENCES}{REFERENCES}}\list
{[\arabic{enumi}]}{\settowidth\labelwidth{[#1]}\leftmargin\labelwidth
\advance\leftmargin\labelsep
\usecounter{enumi}}
\def\newblock{\hskip .11em plus .33em minus -.07em}
\sloppy
\sfcode`\.=1000\relax}
\let\endthebibliography=\endlist

\hoffset = -1truecm
\voffset = -2truecm


\title{\large\bf
Integrable Models And The Toda Lattice Hierarchy
}
\author{
{\normalsize\bf
Bani Mitra Sodermark \thanks{e.mail: bani.sodermark@kau.se}
}\\
\normalsize Dept of Engineering Sciences, Physics and Mathematics, \\
Karlstad University, 65188 Karlstad, SWEDEN}
\date{20 April 1999}
\maketitle

\begin{abstract}

   A pedagogical presentation of integrable models with special reference to
the Toda lattice hierarchy has been attempted. The example of the $KdV$
equation has been studied in detail, beginning with the infinite conserved
quantities and going on to the Lax formalism for the same. We then go on to
symplectic manifolds for which we construct the Lax operator. This formalism
is applied to Toda Lattice systems. The Zakharov Shabat formalism aimed at
encompassing all integrable models is also covered after which the zero
curvature condition and its fallout are discussed. We then take up Toda
Field Theories and their connection to W algebras via the Hamiltonian
reduction of the WZNW model. Finally, we dwell on the connection between
four dimensional Yang Mills theories and the $KdV$ equation along with a
generalization to supersymmetry.

\end{abstract}

\newpage


\section{Introduction: Non-Linear Equations}

Linear partial differential equations, in particular the Schroedinger,
Klein-Gordon and Dirac equations, have been known in field theory over
a long time, and have been used in many different problems with great
success. Non-linear equations, i.e., equations where the potential term
is non-linear in the field ($S$), have been known for some time as well.
These equations and their solutions are the topic of the present Article.
\par
        The earliest non-linear wave equations known in physics were the
Liouville and Sine-Gordon equations. The Liouville equation arose in the
context of a search for a manifold with constant curvature. Pictorially,
such parametrizations may be likened to covering a surface with a fishing
net. Since the knots on the fishing net do not move, the arc length is
constant. The threads in the net correspond to a local coordinate system
on the surface.

\setcounter{equation}{0}
\renewcommand{\theequation}{1.\arabic{equation}}
\par
        The Liouville manifolds may be reparametrized locally so as to
have a metric of the form:
\begin{equation}
A= \left(\matrix {
    \exp{\rho}   &     0 \cr
       0         &  \exp{\rho}
}\right)
\end{equation}
so as to be conformally equivalent to a flat space metric. The study of such
manifolds with constant curvature led J.Liouville [1] to the equation known
by his name:
\begin{equation}
{{\partial}^2 \rho \over {\partial{x}\partial{y}}} = \exp{\rho}
\end{equation}
$x$ and $y$ being local orthogonal coordinates. Interest in this equation
was renewed in the 70's and 80's due to its appearance in string theories
[2,3,4].
\par
        The Sine-Gordon equation, named after a pun on the Klein-Gordon
equation, is an equation for the angle $\omega$ between two coordinate
lines when the total curvature is constant and negative. This equation
first appeared in the work of Enneper in 1870, and has the form:
\begin{equation}
{{\partial}^2 \omega \over {\partial{x}\partial{y}}} = \sin(\omega)
\end{equation}
where $x$ and $y$ are coordinates in a system with constant arc length.
\par
        The Sine-Gordon equation has some interesting solutions known as
{\it solitons} and {\it breathers}. A soliton satisfies three conditions.
First, a single soliton must have constant shape and velocity. Secondly,
it must be localized, and its derivative must vanish at infinity. Thirdly,
if two solutions collide, they should survive the collision with their
shapes unchanged.
\par
        Principally, there are two types of solitons, one which increases
by a fixed amount (say $2\pi$), and is called a `kink'; the other which
decreases by the same amount, and is called an `anti-kink'.
\par
        A breather is a localized solution that varies periodically, and
could be considered as a permanently bound system of a kink and anti-kink.
\par
        An interesting property of the Sine-Gordon equation is that  its
solutions can be mapped into others through the Baecklund transformation
[5], and can thus be used to create new solutions from known solutions.
It is however impossible to generate a complete set of solutions from one
original solution, via the Baecklund transformation [5].
\par
        A third non-linear equation which we shall study in some detail,
was discovered in 1895 by D.J.Korteweg and G. de Vries [6], while trying
to describe the motion of water-waves in a canal. It has the form:
\begin{equation}
u_t - 6uu_x + u_{xxx} = 0
\end{equation}
and is also known as the $KdV$ equation. It has been extensively studied,
and many of the properties of non-linear wave equations that are known
today, were discovered in connection with its solution. This equation was
solved by Gardner, Greene, Kruskal, and Miura in 1967 [7-13]. Along with
N.J. Zabusky and C. H. Su, they also found many interesting properties
of the same. One of these is that the $KdV$ equation has an infinite number
of conservation laws, and that the conserved quantities of each of these
laws can be used as a Hamiltonian for an integrable system. This collection
of Hamiltonians is called the $KdV$ hierarchy.
\par
        There exists a theorem of classical mechanics, which states that if
a Hamiltonian system with $2n$ degrees of freedom has $n$ functionally 
independent conserved quantities such that the Poisson bracket of any two of 
them vanishes, i.e.,the integrals of motion are in `involution', the system 
is completely integrable. It is clear that solutions of systems with an 
infinite number of conserved quantities must be infinitely restricted. A 
soliton is precisely such a solution: it is a localized wave which retains
its shape even after collisions. Intuitively, it is clear that for this to 
happen,there must be an infinite number of conservation laws, and therefore 
an infinite number of conserved quantities. The terms `integrable models' and
`solitons' are often used synonymously.
\par
        A system of coupled equations of motion describing a 1-dimensional
crystal with non-linear coupling between nearest neighbour atoms, was
introduced by M.Toda [14] in 1967. The equations of motion are
\begin{equation}
m {{d^2 r_n} \over {dt^2}} = a[2 e^{-r_n} - e^{-r_{n-1}} - e^{-r_{n+1}}]
\end{equation}
where $r_n = u_{n+1}-u_n$, and $u_n(t)$ is the longitudinal displacement
of the $n$-th atom with mass $m$ from its equilibrium position, $a$ being
a constant. These models admit soliton solutions which have been studied
experimentally on an electrical network by K. Hitota and K. Suzuki [15].
In the continuunm limit, these equations reduce to the $KdV$ equation [5].
\par
        We see that models with exponential interactions are a source of
non-linear equations, the Liouville and Sine-Gordon equations being
examples. The Liouville equation could be generalized to include a mass
term:
\begin{equation}
{{\partial}^2 \phi \over {\partial{x}\partial{y}}} +m^2 \phi = e^\phi
\end{equation}
while the Sine-Gordon equation could be generalized to the "Sinh-Gordon"
equation with the replacement $\omega \rightarrow i\omega$. Thus
\begin{equation}
{{\partial}^2 \omega \over {\partial{x}\partial{y}}} +m^2 \omega
= \sinh \omega
\end{equation}
We also have the Toda Field Theory equations
\begin{equation}
{{\partial}^2 \phi_i \over {\partial{x}\partial{y}}} = -e^{k_{ij}\phi_j}
\end{equation}
Here $k_{ij}$ is the Cartan matrix for some complex Lie Algebra.  The
simplest of these field theories is the $A_r$ Toda field theory, and it
includes the Liouville field theory for the special case $r=1$. There
exist generalizations of the Toda equations called "Affine Toda Equations",
and have an extra term on the RHS, taking the form:
\begin{equation}
{{\partial}^2 \phi_i \over {\partial{x}\partial{y}}} = -e^{k_{ij}\phi_j}
+ \gamma R_i e^{k_0 \phi_j}
\end{equation}
Here $K$ is an affine Cartan matrix, and $R_i$ the right null vector for
this matrix when $R_0$ is normalized to unity.
\par
        These models include the Sinh-Gordon equation as a special case.
Both the Toda and Affine Toda field theories have an infinite number of
conserved quantities [16]. They admit soliton solutions with an imaginary
$\phi_i$ [17]. Both models have been formally solved by Leznov and
Saveliev [18].
\par
        The Toda field theories can be obtained from the Toda Lattice by setting
\begin{equation}
\psi_i = (\phi_i-\phi_{i-1}) - (\phi_{i+1} - \phi_i)
\end{equation}
whence
\begin{equation}
{{\partial}^2 \psi_i \over {\partial{t}^2}} -
{{\partial}^2 \psi_i \over {\partial{x}^2}}
= - [2 e^{\psi_i} - e^{\psi_{i-1}} - e^{\psi_{i+1}}]
\end{equation}
for $SU(n+1)$, showing that the space-independent solutions of (1.11)
satisfy (1.5).
\par
        Since the Toda field theories are the $\gamma=0$ limits of the
Affine Toda field theories, they could be used to classify 2-dimensional
models with a second order phase transition, with the Toda field theory
describing the model at the critical point where it has to be conformally
invariant [19]. Hence the great interest in (Affine) Toda field theories.
However the precise connection is still unclear. Central charges and
critical exponents have been calculated and compared. One hopes that the
Affine Toda field theories are perturbations that correspond to the
physical model away from the critical point. However, more explicit
connections are yet to be found.
\par
        The method originally used for solving non-linear equations, and
especially the $KdV$ equation, was the inverse scattering method
originated by Gelfand and Levitan [20]. This involved looking for a
linear equation related to the original non-linear equation, and studying
the evolution of the latter. In 1968, P.Lax provided this method within
a solid theoretical framework [21]. The Lax equation is
\begin{equation}
L_t + [L, M] = 0
\end{equation}

where $L$ and $M$ are operators satisfying
\begin{equation}
L\psi = \lambda \psi ;
\end{equation}
and
\begin{equation}
\psi_t = M\psi
\end{equation}
where $\lambda$ is a scalar, and $\psi$ a solution of a linear equation
which is just the Schroedinger equation for the $KdV$ case ! The Lax
equation was generalized to the form of a zero curvature condition which
facilitates greatly the form of the transition matrix from the initial
to the final state.
\par
        In what follows, we attempt to give a pedagogical presentation
of Integrable Systems with special emphasis on the $KdV$ and Toda systems.
After an introduction to the $KdV$ equation and its properties, we show
how an infinite number of conserved quantities arise via the Muira [8]
transformation, while detailed calculations are referred to ref.[22].
We then dwell on solutions of the $KdV$ equation via the inverse scattering
method and the Lax formalism [21], after which we obtain the Lax operator
for symplectic manifolds, using the Toda Lattice as an example.
The group structure of the Toda equations for $SU(N)$ is also studied.
The Lax transformation was later generalized by Zakharov and Shabat [23]
to a first order formalism which was used by Ablowitz, Kamp, Newell and
Segur (AKNS) [24], for a unified description of other integrable models.
The essential features of this approach are also discussed. A fall-out
of the above is the `zero curvature condition' that facilitates the
transition to the quantum case. However, the treatment we follow is
strictly classical.
\par
        Next we take up the Toda field theories, and after reporting
briefly the connection with conformal invariance, dwell on the Hamiltonian
reduction of the WZNW model to the Toda field theory, which in
effect transforms an affine Lie Algebra to a W-Algebra. (Most calculational
details are skipped, but may be found in the literature [25]). Finally we
refer to the interesting connection between the 4D self-dual Yang-Mills
theory and 2D Integrable models, and the generalization to SuperSymmetry.
\par
        The material is presented as follows. In Sect.2, we introduce the
$KdV$ equation and its conserved quantities. In Sect.3, solutions of
non-linear equations are taken up, in particular the inverse scattering
method and the Lax formalism. In Sect.4, we digress to Symplectic Manifolds
and construct conserved quantities for these manifolds. Sect.5 applies the
above framework to the Toda Lattice where the group structure of the Toda
equations is also discussed. In Sect.6, we take up the unifying first order
formalism of Zakharov and Shabat [23], continuing in Sect.7 to the zero
curvature formalism and its ramifications. In Sect.8, we take up Conformal
Invariance, and introduce Toda Field Theories which are constructed
independently of the Toda Lattice. In Sect.9, we carry out the Hamiltonian
reduction of the WZNW model to Toda Field Theories. Finally in Sect.10,
we take up the connection of Toda Field Theories with Self-dual Yang-Mills
models. Sect.11 contains some concluding remarks.

\section{The $KdV$ Equation}

The $KdV$ equation was formulated to explain the solitary water waves
observed
by J.Scott Russell in the Edinburgh Glassgow canal. It is a non-linear
equation in one space and one time dimension and possesses soliton
solutions.
Of this, however, nothing was known at the time of its formation.

\setcounter{equation}{0}
\renewcommand{\theequation}{2.\arabic{equation}}

\par
        The $KdV$ equation after an initial scaling takes the form
\begin{equation}
{\partial u \over {\partial t}} = u {\partial u \over {\partial x}}
+{\partial^3 u \over {\partial x^3}}
\end{equation}

        This equation is Galilean invariant, but not Lorentz invariant.
It can be derived from the Hamiltonian
\begin{equation}
H(u) = \int_{-\infty}^{+\infty} [{u^3 \over 6} - {1 \over 2}
({\partial u \over {\partial x}})^2] dx
\end{equation}
where the $u(x)$ satisfy the Poisson bracket relations
\begin{equation}
[u(x),u(y)]= \partial_x \delta (x-y)
\end{equation}
However, the Lagrangian from which it can be derived, is non local:
\begin{equation}
L_{KdV}= {1\over 2} \int_{-\infty}^{+\infty} dx dy u(x) \epsilon(x-y)
{\partial u(y) \over {\partial t}} - \int dx [{u^3 \over 6} - {1 \over 2}
({\partial u \over {\partial x}})^2]
\end{equation}
where
\begin{equation}
\epsilon(x-y)=\theta(x-y)-{1\over 2},
\end{equation}
$\theta$ being the step function.  Ergo, one cannot write down a local
Lagrangian whose Euler-Lagrange equations yield the $KdV$ equation.
\par
        Solutions of the $KdV$ equation can be shown to be soliton
solutions which travel without any change of shape. It is the non-linear
term which is responsible for the above property.
\par
        What is most interesting about the $KdV$ equation is that it
admits of an infinite number of conserved quantities as was shown by
Miura [8]. This procedure is explained below.
\par
        The $KdV$ equation is related to another equation called the
modified $KdV$ ($MKdV$) equation, viz.,
\begin{equation}
{\partial v \over {\partial t}} = v^2 {\partial v \over {\partial x}}
+{\partial^3 v \over {\partial x^3}}
\end{equation}
where $v$ is related to $u$ in the $KdV$ equation through the
Riccati transformation
\begin{equation}
u = v^2 + i {\sqrt 6} {\partial v \over {\partial x}}
\end{equation}
        The $MKdV$ equation is however not Galiliean invariant.
Under the transformation
\begin{equation}
t \rightarrow t ; \quad x \rightarrow x + {3t \over {2\epsilon^2}}; \quad
u \rightarrow u + {3 \over {2\epsilon^2}}; \quad v \rightarrow
{v\epsilon \over {\sqrt 6}} + {{\sqrt 6} \over {2\epsilon}}
\end{equation}
it reduces to
\begin{equation}
\partial_t v = ({\epsilon^2 v^2 \over 6} + v) {\partial_x v}
+ \partial_x^3 v  \\
= \partial_x [{{\epsilon^2 v^3} \over 18} + {v^2 \over 2 }+ \partial_x^2 v]
\end{equation}
This yields a solution of the $KdV$ equation through the transformation
\begin{equation}
u = \epsilon^2 v^2/6 + v + i\epsilon \partial_x v
\end{equation}
The second form of (2.9) is in the nature of a continuity equation, so that
we can identify
\begin{equation}
K= \int_{-\infty}^{+\infty} dx v(x(t))
\end{equation}
as the conserved quantities. $v$ can be inverted in terms of $u$ as
\begin{equation}
v = \sum_{0}^{\infty} \epsilon^n v_n(u(x,t))
\end{equation}
and this yields $v_n(u(x,t))$  as the conserved densities, since each power
of $\epsilon$ must independently satisfy a continuity equation. That these
are also in involution can also be checked, being explicitly shown by Das
[22]. Some of the conserved quantities are
\begin{equation}
v_1=-i\partial_x u_1; \quad v_2=-{u^2 \over 6}-\partial_x^2 u; \quad
v_3=i\partial_x [{u^2 \over 3}+\partial_x^2 u]
\end{equation}

\section{The Lax Framework}
\setcounter{equation}{0}
\renewcommand{\theequation}{3.\arabic{equation}}

Linear Hamiltonian systems with fixed initial value problems can be
solved using the Laplace or Fourier transformations. Such methods are
inapplicable for the nonlinear equations and new methods must be found.
Gardner, Green, Krushal and Miura [9] managed to solve the initial value
problem for the $KdV$ equation in a very ingenious way. In subsequent years,
this method has become the standard method for solving non-linear systems
and
goes by the name of inverse scattering theory [20,21]. This method is
outlined in Fig 1.

\begin{figure}[t]

\caption{}
\vspace{0.5in}

\begin{picture}(450,175)(-30,-10)

\put(70,120){\framebox(50,20){$u(x,0)$}}
\put(70,20){\framebox(50,20){$u(x,t)$}}

\multiput(95,40)(0,5){16}{\line(0,1){3}}

\put (95,80){\line(-1,1){10}}
\put (95,80){\line(1,1){10}}

\put (120,130){\line(1,0){109}}
\put (120,30){\line(1,0){109}}

\put (174,136.5){\line(3,-2){10}}
\put (174,123){\line(3,2){10}}

\put (174,36.5){\line(-3,-2){10}}
\put (174,23){\line(-3,2){10}}

\put(230,120){\framebox(100,20){Scattering data}}
\put(230,20){\framebox(160,20){Time evolved scattering data}}

\put (280,40){\line(0,1){80}}

\put (280,80){\line(-1,1){10}}
\put (280,80){\line(1,1){10}}

\end{picture}
\end{figure}

The initial value for the partial differential equation is used as the
potential in a 1-dimensional scattering problem for a linear equation,
e.g. the Schroedinger equation. One then finds the so called scattering
data,
i.e. discrete spectrum, normalization constants, reflection constants (as a
function of the wave number) for this scattering problem. Using the partial
differential equation ($pde$) evaluated for $|{x}|$ asymptotically large,
(and hence the $pde$ becomes a linear equation because the potential is
assumed to vanish at spatial infinity), the values of the scattering data
can be found for all later times. Finally, the scattering data allow one
to reconstruct the potential, and hence the solution of the $pde$ for any
later time.
\par
        One would intuitively like a better understanding of the origin
and relevance of the linear Schroedinger equation. One way to see this
is through a generalized Riccati relation of the form:
\begin{equation}
u + 6\lambda = v^2 + i {\sqrt 6} {{\partial v} \over {\partial x}}
\end{equation}
so that the $KdV$ relation (2.1) reduces to
\begin{equation}
{\partial v \over {\partial t}} - (v^2-6\lambda)
{\partial v \over {\partial x}} - {\partial^3 v \over {\partial x^3}} = 0
\end{equation}
\par
        As mentioned earlier, a solution of the $MKdV$ equation yields a
solution of the $KdV$ equation through the Riccati relation. The simplest
way to attempt an inversion of the Riccati relation is to linearize it. To
that end we define
\begin{equation}
v = i {\sqrt 6} \psi_x /\psi
\end{equation}
so that (3.1) takes the form
\begin{equation}
u + 6\lambda = - 6 \psi_{xx}/\psi,
\end{equation}
or equivalently,
\begin{equation}
\psi_{xx} + ({u \over 6} + \lambda) \psi = 0
\end{equation}
which is, in fact, the time-independent Schroedinger equation. There exists
however a more formal theory due to Lax [21], which we now elaborate.
\par
        Given a linear equation described by a time-independent Hamiltonian
$H$, and an operator $A$ whose expectation values are time independent,
$A(t)$ is unitarily equivalent to $A(0)$:
\begin{equation}
U^\dagger(t) A(t) U(t) = A(0)
\end{equation}
where $U(t)$ is the time-evolution operator with the form
\begin{equation}
U(t) = \exp [-iHt]
\end{equation}
Differentiating (3.6) gives
\begin{equation}
U^{\dagger}(t)({{\partial A} \over {\partial t}} - i[A,H]) U(t) = 0
\nonumber
\end{equation}
which implies
\begin{equation}
{{\partial A} \over {\partial t}} = i[A,H]
\end{equation}
Thus for the expectation value of $A(t)$ to be time independent, the
standard time evolution relation (3.8) must be satisfied. Further, from
eq.(3.7) follows the relation
\begin{equation}
{{\partial U(t)} \over {\partial t}} = -i H U(t) = B U(t)
\end{equation}
where
\begin{equation}
B = - i H
\end{equation}
is an anti-Hermitian operator.
\par
        This argument is mimicked in the case of a non-linear evolution
equation. Let
\begin{equation}
L(u(x,t)) = L(t)
\end{equation}
denote the linear operator we seek. We assume it to be Hermitian, and to
have eigen-values independent of $t$. For this to be true, one must have
$u^\dagger(t)L(t)u(t)$$=$$L(0)$.
Differentiating both sides w.r.t. $t$, we obtain
\begin{equation}
{{\partial U^\dagger(t)} \over {\partial t}} L(t) + U^\dagger(t)
{{\partial L(t)} \over {\partial t}} U(t) +U^\dagger(t)L(t)
{{\partial U(t)} \over {\partial t}}
\end{equation}
Unlike the linear case, we do not know the form of $U(t)$. However, $U$
is unitary, so
\begin{equation}
U^\dagger U=1 \Rightarrow  {{\partial U^\dagger(t)} \over {\partial t}} U(t)
+ U^\dagger {{\partial U(t)} \over {\partial t}}=0
\end{equation}
Thus we can write
\begin{equation}
{{\partial U(t)} \over {\partial t}} = B(t) U(t)
\end{equation}
where anti-hermiticity must be imposed on $B$. Substitution in (3.12), and
a little simplification, yields
\begin{equation}
{{\partial L(t)} \over {\partial t}} = [B(t), L(t)]
\end{equation}
which is similar to (3.8), except for the fact that we do not yet know the
form of $B$. However, let us assume that $L(t)$ is linear in $u(x,t)$.
Consequently, the LHS of (3.14) is a multiplicative operator, proportional
to the time evolution operator of $u(x,t)$. This would ensure that the
eigen-values $\lambda$ of $L(t)$ would be time-independent, i.e.,
\begin{equation}
L(t) \psi(t) = - \lambda \psi(t)
\end{equation}
Further, $\psi(t)$ must be unitarily related to its value at $t=0$, i.e.,
\begin{equation}
\psi(t) = U(t) \psi(0),
\end{equation}
and its evolution w.r.t. time would take the form
\begin{equation}
{{\partial \psi(t)} \over {\partial t}} =
{{\partial U(t)} \over {\partial t}} \psi(0) = B(t) \psi(t)
\end{equation}
The operators $L(t)$ and $B(t)$, when they exist, are known as the Lax
pair, corresponding to a given non-linear evolution equation, and play
a fundamental role in determining the solution. For the $KdV$ equation,
$L(t)$ is obtained from the linear form of the Schroedinger equation
\begin{equation}
L(t) = D^2 + {1 \over 6} u(x,t);  \quad D \equiv {\partial \over {\partial x}}.
\end{equation}
By trial and error, $B(t)$ can be chosen so that (3.15) is satisfied,
and a possible solution is
\begin{equation}
B(t) = 4D^3 + {1 \over 2} (Du + uD)
\end{equation}
The solution for $\psi$ w.r.t. $t$ follows from (3.18) and (3.20) to be
\begin{equation}
\psi_t = 4\psi_{xxx} +{1 \over 2}u_x \psi + u \psi_x + const. \psi
\end{equation}
which yields, using the Schroedinger equation (3.5):
\begin{equation}
\psi_t + {1 \over 6}u_x \psi - {1 \over 3}u \psi_x + 4 \lambda \psi_x = const. \psi
\end{equation}
\par
        A.Lenard [26], in an unpublished report, further displayed the
relation between the Schroedinger equation and the $KdV$ relation by
elegantly deriving the latter from the former, using only the assumption
that the spectral parameter $\lambda$ in (3.4) is time-independent.
\par
        The $KdV$ equation exhibits also a fascinating symmetry, i.e.,
that of the group $SL(2,R)$. Consider a group element
\begin{equation}
g = \exp [i \theta^a T_a]
\end{equation}
where $T_a$ is a generator of $SL(2,R)$, and define
\begin{equation}
A_\mu \equiv g^{-1} {\partial_\mu} g
\end{equation}
Then the $KdV$ equation follows from the fact that the Maurer-Cartan
equation
\begin{equation}
{\partial_\mu}A_\nu - {\partial_\nu}A_\mu - [A_\mu, A_\nu] = 0
\end{equation}
is satisfied for a special for a special choice of gauge, e.g.,
\begin{equation}
A^1_1 = - {\sqrt \lambda}; (\lambda < 0); \quad
A^3_1 = 6; \quad A^2_1 = -{1 \over 36} u(x,t); \quad A^3_0 =A(u(x,t))
\end{equation}

\section{Lax Formalism On Symplectic Manifolds}
\setcounter{equation}{0}
\renewcommand{\theequation}{4.\arabic{equation}}

In this Section, we conclude the above study of the $KdV$ equation with a
with a short discussion on symplectic geometry, which is directly relevant
for application to the Toda Lattice.
\par
        A symplectic manifold is one with a preferred 2-form
$f_{\mu\nu}$ which is non-degenerate and closed. The phase space of an
integrable model corresponds to a very special symplectic manifold, since
it possesses a dual Poisson bracket structure. We assume that there exist
two distinct 2-forms which are both non-degenerate and closed. One way of
expressing the existence of two distinct symplectic structures is to
require that the same dynamical equation be described by two distinct
first order Lagrangians $L_0$ and $L$, where
\begin{equation}
L_0 = \theta_\mu ^{(0)}(y) {\dot y}^\mu - H_0(y);
\end{equation}
\begin{equation}
L = \theta_\mu(y) {\dot y}^\mu - H(y)
\end{equation}
where
\begin{equation}
{\dot y}^\mu = {{dy^\mu} \over {dt}}; \quad [\mu = 1,2,...2N]
\end{equation}
The Euler-Lagrangian equations following from (4.1-2) are
\begin{equation}
f_{\mu\nu}(y) y^\nu = \partial_\mu H_0(y)
\end{equation}
\begin{equation}
F_{\mu\nu}(y) y^\nu = \partial_\mu H(y)
\end{equation}
where
\begin{equation}
f_{\mu\nu} = \partial_\mu \theta_\nu^{(0)}(y) -
\partial_\nu \theta_\mu^{(0)}(y)
\end{equation}
\begin{equation}
F_{\mu\nu} = \partial_\mu \theta_\nu(y) - \partial_\nu \theta_\mu(y)
\end{equation}
\par
        It is easy to see that the two forms $f$ and $F$ are closed, where
\begin{equation}
f= {1 \over 2} f_{\mu\nu} dy^\mu \wedge dy^\nu
\end{equation}
\begin{equation}
F= {1 \over 2} F_{\mu\nu} dy^\mu \wedge dy^\nu
\end{equation}
since $f_{\mu\nu}$ and $F_{\mu\nu}$ satisfy the Bianchi identities
\begin{equation}
{\partial_\lambda} f_{\mu\nu} + {\partial_\mu} f_{\nu\lambda}
+ {\partial_\nu} f_{\lambda\mu} = 0;
\end{equation}
and
\begin{equation}
{\partial_\lambda} F_{\mu\nu} + {\partial_\mu} F_{\nu\lambda}
+ {\partial_\nu} F_{\lambda\mu} = 0.
\end{equation}
Besides, they must also be non-degenerate since (4.4) and (4.5) describe
the same dynamical system. Let their universes be $f^{\mu\nu}$ and
$F^{\mu\nu}$, i.e.,
\begin{equation}
f_{\mu\nu} f^{\nu\lambda} = F_{\mu\nu} F^{\nu\lambda} = \delta_\mu^\lambda
\end{equation}
so that (4.4-5) take the forms
\begin{equation}
{\dot y}^\nu = f^{\nu\mu} \partial_\mu H_0 (y)
\end{equation}
\begin{equation}
{\dot y}^\nu = F^{\nu\mu} \partial_\mu H (y).
\end{equation}
We can also construct a nontrivial $(1,1)$ tensor $S_\mu^\nu$ as
\begin{equation}
S_\mu^\nu = F_{\mu\lambda}(y) f^{\lambda\nu}(y).
\end{equation}
Consistency of (4.4) and (4.5) further requires that
\begin{equation}
\partial_\mu\partial_\nu H_0(y) - \partial_\nu\partial_\mu H_0(y) = 0
\end{equation}
so that after a little algebra, one can show that
\begin{equation}
{{df_{\mu\nu}(y)} \over {dt}} = -U_\mu^\lambda f_{\lambda\nu}
+ U_\nu^\lambda f_{\lambda\mu}
\end{equation}
where
\begin{equation}
U_\mu^\nu = \partial_\mu y^\nu = \partial_\mu [f^{\nu\lambda}
\partial_\lambda H_0(y)]   \\  \nonumber
= \partial_\mu [F^{\nu\lambda} \partial_\lambda H(y)]
\end{equation}
with a corresponding relation for $F^{\mu\nu}$, i.e.,
\begin{equation}
{{dF_{\mu\nu}(y)} \over {dt}} = -U_\mu^\lambda F_{\lambda\nu}
+ U_\nu^\lambda F_{\lambda\mu}
\end{equation}
involving the same $U$-tensor. The corresponding equations for the
inverses $f^{\mu\nu}$ and $F^{\mu\nu}$ follow from (4.17) and (4.19),
and have the forms
\begin{equation}
{{df^{\mu\nu}} \over {dt}} = f^{\mu\lambda} U_\lambda^\nu -
f^{\nu\lambda} U_\lambda^\mu
\end{equation}
\begin{equation}
{{dF^{\mu\nu}} \over {dt}} = F^{\mu\lambda} U_\lambda^\nu -
F^{\nu\lambda} U_\lambda^\mu
\end{equation}
We can finally show that
\begin{equation}
{{dS_\mu^\nu} \over {dt}} = S_\mu^\lambda U_\lambda^\nu -
U_\mu^\lambda S_\lambda^\nu
\end{equation}
which in matrix notation
\begin{equation}
{{dS} \over {dt}} = [S,U]
\end{equation}
can be recognized as a Lax equation (3.15), thus providing a Lax
representation of the dynamical equations (4.13) and (4.14). One
important consequence of (4.23) is that the set of quantities
\begin{equation}
K_n = {1 \over n} Tr S^n
\end{equation}
and
\begin{equation}
K_0 = ln {\mid det S \mid}
\end{equation}
can be shown to be invariants since
\begin{equation}
{{dK_n} \over {dt}} = Tr[P(S) {dS \over dt}] = Tr[P(S)[S,U]] = 0
\end{equation}
$P(S)$ is a polynomial in $S$. That these are in involution can easily be
checked, as done explicitly in ref.[22]. Applied to the $KdV$ equation,
the two Poisson structures of that equation are given by the correspondence:
\begin{equation}
F^{\mu\nu} \rightarrow D ;
\end{equation}
\begin{equation}
f^{\mu\nu} \rightarrow D^3 + {1 \over 3} (Du + uD)
\end{equation}
Going to the coordinate bases we have
\begin{equation}
F(x,y) = <y \mid D \mid x> = \partial_\lambda \delta(x-y)
\end{equation}
\begin{equation}
f(x,y) = {{\partial^3} \over {{\partial x}^3}} + {1 \over 3}
(\partial_x u + u \partial_x) \delta(x-y)
\end{equation}
so that
\begin{equation}
F^{-1}(x,y) = \epsilon(x-y) = \theta(x-y) - {1 \over 2}
\end{equation}
However $f^{-1}(x-y)$ cannot be expressed in a closed form. The Lax
operator $S$ takes the form
\begin{equation}
S = D^2 + {2 \over 3}u + {1 \over 3}(Du)D^{-1},
\end{equation}
and with a little algebra, (4.23) can be shown to be reduced to the
$KdV$ equation, with consequently an infinite $\#$ of conserved
quantities. This is described in detail in ref.[22].

\section{The Toda Lattice}
\setcounter{equation}{0}
\renewcommand{\theequation}{5.\arabic{equation}}

The model of the $KdV$ equation that has been studied so far is a
continuum model. A finite dimensional system with a finite $\#$ of
degrees of freedom is simpler to study. The Toda Lattice is such a
system to which the symplectic approach of the above Section  is
especially applicable. We now study the Toda Lattice and its
integrability from a symplectic point of view, following it up with
a group theoretical treatment.
\par
        The Toda Lattice describes the motion of $N$ point masses on
the line, under the influence of an exponential interaction. The
Hamiltonian equations in terms of the canonical coordinates $Q_i$ and
momenta $P_i$ are given by
\begin{eqnarray}
{\dot Q}_i  &=& P_i; \quad (i= 1,2,....N); \\  \nonumber
{\dot P}_j  &=& e^{-(Q_j - Q_{j-1})} - e^{-(Q_{j+1} - Q_j)};
   \quad (j=2,3,.N-1);  \\  \nonumber
{\dot P}_1  &=& - e^{-(Q_2-Q_1)}; \quad  {\dot P}_N = e^{-(Q_N-Q_{N-1})}.
\end{eqnarray}
The equations can be cast into a more symmetrical form by enlarging the
system to $(N+2)$ point masses, with end points at spatial infinity. In
that case, the Hamiltonian equations take the form :
\begin{equation}
{\dot Q}_i = P_i; \quad (i= 1,2,....N); \\  \nonumber
{\dot P}_i = e^{-(Q_i - Q_{i1})} - e^{-(Q_{i1} - Q_i}.
\end{equation}
We can choose
\begin{equation}
y^i = Q_i ; \quad y^{N+i} = P_i; \quad (i=1,2,.. N).
\end{equation}
Applying the geometrical method of the previous Section, two choices of
the Lagrangian are as follows:
\begin{equation}
L_0=\sum_{i=1}^{N} [{1 \over 2}(P_i{\dot Q}_i-Q_i{\dot P}_i)-{1 \over
2}P_i^2
+{e^{-(Q_{i+1}-Q_i)}}];
\end{equation}
\begin{equation}
L= \sum_{i=1}^{N} [{1 \over 2}(P_i^2+{e^{-(Q_{i+1}-Q_i)}}){\dot Q}_i +
\pi_i(P){\dot P}_i] -H(Q,P)
\end{equation}
where
\begin{equation}
\pi_i(P) = {1 \over 2} \sum_{j=1}^{N} \epsilon(i-j){\dot P}_j;
\end{equation}
\begin{equation}
H(Q,P) = \sum_{i=1}^{N} [{P_i^3 \over 3} +
(P_i+P_{i+1}){e^{-(Q_{i+1}-Q_i)}}]
\end{equation}
$f_{\mu\nu}$ turns out to have the canonical Poisson bracket structure
\begin{equation}
f_{\mu\nu} = \left(\matrix {
       0   &   -I \cr
       I   &    0
}\right)
\end{equation}
so that
\begin{equation}
f^{\mu\nu} = \left(\matrix {
        0   &    I \cr
       -I   &    0
}\right)
\end{equation}
$F_{\mu\nu}$ can be shown to have the form [22]
\begin{equation}
F_{\mu\nu} = \left(\matrix {
       A   &   B \cr
       B   &   e
}\right)
\end{equation}
where
\begin{eqnarray}
A_{ij}  &=& \delta_{i+1,j} {e^{-(Q_{i+1}-Q_i)}} -
\delta_{i,j+1} {e^{-(Q_{j+1}-Q_j)}}   \\  \nonumber
B_{ij}  &=& P_i \delta_{ij};  \quad e_{ij} = \epsilon(j-i)
\end{eqnarray}
The $(1,1)$ tensor $S_\mu^\nu$ thus takes the form
\begin{equation}
S_\mu^\nu = \left(\matrix {
       B   &   A \cr
      -e   &   B
}\right)
\end{equation}
and the conserved quantities are
\begin{equation}
Tr S =2 Tr B = 2 \sum_{i=1}^{N} P_i;
\end{equation}
\begin{equation}
{1 \over 2} Tr S^2 = Tr [2B^2-(Ae+eA)]   \\  \nonumber
= \sum_{i=1}^{N} [{P_i^2 \over 2} + {e^{-(Q_{i+1}-Q_i)}}] \equiv H_0(Q,P);
\end{equation}
\begin{equation}
{1 \over 6}Tr S^3 = \sum_{i=1}^{N}
[{P_i^3 \over 3} + (P_i+P_{i+1}){e^{-(Q_{i+1}-Q_i)}}] \equiv H(Q,P)
\end{equation}
The Lax representation (4.23) for the Toda equation takes the form
of the following matrix equations
\begin{equation}
{dA \over {dt}} = -[B,D];
\end{equation}
\begin{equation}
{dB \over {dt}} = A -De = {1 \over 2} [e,D]
\end{equation}
which reduce to the Toda equations ${\dot Q}_i = P_i$ and
${\dot P}_i$ = $e^{-(Q_i-Q_{i-1})} -e^{-(Q_{i+1}-Q_i)}$ respectively.

\subsection{Group Structure of Toda Equations}

Eq.(5.1) can be differentiated and put in the form
\begin{eqnarray}
{\ddot Q}_1  &=& -e^{-(Q_2-Q_1)}  \\   \nonumber
{\ddot Q}_i  &=& {\dot P}_i = e^{-(Q_i-Q_{i-1})} - e^{-(Q_{i+1}-Q_i)} \\
\nonumber
{\ddot Q}_N  &=& {\dot P}_N = e^{-(Q_N-Q_{N-1})}
\end{eqnarray}
It is easily checked that
\begin{equation}
\sum_{i=1}^{N} {\ddot Q}_i = \sum_{i=1}^{N} {\dot P}_i = 0
\end{equation}
i.e., the total momentum is conserved, and therefore the centre of mass
motion can be separated and the dynamics of the system expressed in terms
of $(N-1)$ coordinates and momenta. Defining
\begin{equation}
q_a = Q_{a+1} - Q_a;  \quad a = {1,2...N-1},
\end{equation}
the second order equations satisfied by the $q_a$'s can be written as
\begin{eqnarray}
{\ddot q}_1  &=& 2 e^{-q_1} - e^{-q_2} \\   \nonumber
{\ddot q}_a  &=& -e^{-q_{a-1}} +2 e^{-q_a} -e^{-q_{a+1}}; \quad a=1,..N-1
\\  \nonumber
{\ddot q}_N  &=& -e^{-q_{N-1}} + 2 e^{-q_N}
\end{eqnarray}
which can be compactly written as
\begin{equation}
{\ddot q}_a = \sum_{b=1}^{N-1} K_{ab} e^{-q_b}
\end{equation}
$K_{ab}$ being the Cartan matrix for $SU(N)$. Eq.(5.22) generalizes for
the other Lie Algebras as well.
\par
        The Lagrangian giving rise to the above Euler-Lagrangian equations
can be written as
\begin{equation}
L = \sum_{a=1}^{N} \sum_{b=1}^{N} {1 \over 2} {\dot q}_a {K_{ab}}^{-1} {\dot q}_b
-\sum_{a=1}^{N} e^{-q_a}
\end{equation}
${K_{ab}}^{-1}$ being the inverse of the Cartan matrix. The momenta
conjugate to $q_a$ are defined as
\begin{equation}
p_a ={{\partial L} \over {\partial {\dot q}_a}}
= \sum_{b=1}^{N-1} {K_{ab}}^{-1} {\dot q}_b
\end{equation}
and it is easily checked that
\begin{equation}
\{q_a, p_b \} = \delta_{ab}
\end{equation}
so that $\{q_a,p_a\}$ constitute a canonical coordinate system.
\par
        That the group structure entering above is not just accidental,
can be seen by defining the following Lax operators:
\begin{equation}
S = {1 \over 2} \sum_{a=1}^{N}[p_aH_a + (E_a+E_{-a}) {e^{-q_a/2}}];
\end{equation}
\begin{equation}
U = -{1 \over 2} \sum_{a=1}^{N-1} {e^{-q_a/2}}[E_a - E_{-a}]
\end{equation}
where $H_a$ and $E_a$ are the generators of $SU(N)$ in the Chevally basis.
\par
        The Lax equation (4.23) can be seen to be satisfied, since
${{dS} \over {dt}}-[S,U]$ reduces to
$$ {1 \over 2} \sum_{a,b=1}^{N-1} H_a {K_{ab}}^{-1}
[{\ddot q}_b - \sum_{c=1}^{N-1} K_{bc}{e^{-q_c}}] $$
which is zero by virtue of the Toda equations (5.22). Hence the quantities
\begin{equation}
K_n = {1 \over n} Tr S^n
\end{equation}
must be conserved under the flow of the Toda equations. Since $S$ belongs
to the $SU(N)$ algebra, the number of independent conserved quantities can
equal $(N-1)$, which is the rank of $SU(N)$. The total number of conserved
quantities is thus $N$, if we add the total momentum. It can be shown that
these are also in involution [22]. This treatment is due to Leznov and
Saveliev [18].

\section{Zakharov-Shabat Formalism}
\setcounter{equation}{0}
\renewcommand{\theequation}{6.\arabic{equation}}

So far we have only studied two integrable models, viz., the continuum
$KdV$ and the finite dimensional Toda Lattice. In trying to understand
the non-linear Schroedinger equation which is also integrable, Zakharov
and Shabat [23] obtained a description which was later generalized by
AKNS [24] to describe various other integrable models. This approach uses
a Lax operator which is first order in the derivative $\partial_x$, in
contrast to the second order formalism in eq.(3.19). Besides describing
various integrable models in a unified manner, this approach has the
additional advantage that the inverse scattering method generalizes
readily to the quantum case. In what follows, we describe the first order
formulation of the Lax operator, and elucidate the essential features of
this approach. It is easily checked that if
\begin{equation}
L(t) \psi(t) = -\lambda \psi(t);
\end{equation}
\begin{equation}
\partial_t L(t) = [B(t), L(t)],
\end{equation}
where
\begin{equation}
{{\partial \psi(t)} \over {\partial t}} = B(t) \psi(t),
\end{equation}
then
\begin{equation}
{{\partial \lambda(t)} \over {\partial t}} = 0.
\end{equation}
We can invert the argument to identify the Lax pair in the following way.
Namely, if
$$L(t) \psi(t) = -\lambda \psi(t);
{{\partial \psi(t)} \over {\partial t}} = B(t) \psi(t),
with  {{\partial \lambda(t)} \over {\partial t}} = 0 $$,
i.e., if the compatibility condition of (6.1) and (6.3) yield the system
under study, then $L(t)$ and $B(t)$ can be identified as the Lax pair of
the system. We would like $L(t)$ to be linear in $\partial_x$. Using the
analogy between the Klein-Gordon and Dirac equations, we define a
two-component column matrix
\begin{equation}
\phi = \left(\matrix {
       \phi_1 \cr
       \phi_2
}\right)
\end{equation}
and generalize the two equations (6.1) and (6.3) to first order
matrix equations as
\begin{eqnarray}
{{\partial \phi} \over {\partial x}} &=& (q\sigma_+ + r \sigma_
-i\rho \sigma_3) \phi;  \\  \nonumber
{{\partial \phi} \over {\partial t}} &=& (P\sigma_+ Q \sigma_
+ R \sigma_3)\phi
\end{eqnarray}
where $\sigma_{\pm}$ and $\sigma_3$ are the Pauli spin matrices. The
dynamical variables $q(x,t)$ and $r(x,t)$ do not depend on the spectral
parameter $\rho$ which is assumed to be independent of $x$ and $t$. The
coefficient functions $P$, $Q$ and $R$ on the other hand, do depend on
$\rho$, and are functionals of $q$ and $r$. Demanding that the partial
derivatives of $\phi$ w.r.t. $x$ and $t$ commute, we obtain the
compatibility conditions to be
\begin{equation}
{{\partial R} \over {\partial x}} = qQ - rP;
\end{equation}
\begin{equation}
{{\partial r} \over {\partial t}} = {{\partial Q} \over {\partial x}}
-2 r R -2i \rho Q ;
\end{equation}
\begin{equation}
{{\partial q} \over {\partial t}} = {{\partial P} \over {\partial x}}
+ 2 q R + 2i \rho P
\end{equation}
i.e., if (6.7-9) describe the non-linear evolution of a system, then
(6.6) describes the Lax pair appropriate for such a system. Explicitly
\begin{equation}
L = \partial_\lambda - q \sigma_+ - r \sigma_-
\end{equation}
\begin{equation}
B = P \sigma_+ + Q \sigma_- + R \sigma_3
\end{equation}
so that (6.2) is satisfied.
\par
        The choice of $r=6$ yields the $KdV$ equation, and the choice
$r-q=-iv/{\sqrt 6}$, the $MKdV$ equation. The choice $q= {\sqrt k} \psi^*$
and $r= {\sqrt k}\psi$, $k$ being an arbitrary constant parameter, yields
the non-linear Schroedinger equation:
\begin{equation}
i{\partial_t}\psi = -\psi_{xx} + 2 k {\mid \psi \mid}^2 \psi
\end{equation}
and the choice
$r = -q = {1 \over 2}\omega_x$ with
$$ P = Q = {i \over {4\rho}} {\sin \omega}$$
yields the sine-Gordon equation.
\par
        The operator $(L+\lambda)$ in (6.1) can be rewritten as
$v(x,t,\lambda)+\partial_x$, where
\begin{equation}
v = -q\sigma_+ -r\sigma_3 +i\rho \sigma_3
\end{equation}
If one knows the solution of the associated Schroedinger equation at
some other point $(x,t)$ by multiplying the solution by a hermitian
matrix $T(x,y,t,\lambda)$, i.e.,
\begin{equation}
\psi(x,t,\lambda) = T(x,y,t,\lambda)\psi(y,t,\lambda)
\end{equation}
where $T(x,y,t,\lambda)$ is a solution of
\begin{equation}
\partial_x T(x,y,t,\lambda) = -(q\sigma_+ - r\sigma_- +i\rho \sigma_3)
T(x,y,t,\lambda)
\end{equation}
with the initial condition $T(x,x,t,\lambda)=I$.

\section{The Zero Curvature Condition}
\setcounter{equation}{0}
\renewcommand{\theequation}{7.\arabic{equation}}

The Lax condition (6.2) can be written as
\begin{equation}
[\partial_t -B, L] = 0
\end{equation}
Using
\begin{equation}
L = \partial_x -A(x)
\end{equation}
we obtain the form
\begin{equation}
[(\partial_t - B), (\partial_x - A)] = 0
\end{equation}
which is like a zero-curvature condition for
\begin{equation}
F_{01} = [(\partial_0 - A_0), (\partial_1 - A_1)]
\end{equation}
with the identification
\begin{equation}
A_0 = -B(x,\rho); A_1 = -A(x,\rho)
\end{equation}
The importance of the zero curvature condition stems from the fact that
(6.6) may be solved, using
\begin{equation}
\psi(x) = T(x,y, \rho) \psi(y)
\end{equation}
where the transformation
\begin{equation}
T(x,y,\rho) = P_r \exp [- \int_{y}^{x} A_1(z) dz]
\end{equation}
where $P_r$ denotes path ordering.
\par
        It is easy to see that $T(x,y,\rho)$ translates solutions of
the problem along the $x$-axis for a fixed time, i.e.,
\begin{equation}
T(x,y,\rho) T(y,z,\rho) = T(x,z,\rho);
\end{equation}
\begin{equation}
T^{-1}(x,y,\rho) = T(y,x,\rho);
\end{equation}
\begin{equation}
T(x,x,\rho) = 1
\end{equation}
Setting
\begin{equation}
U_r(x_2,t_2; x_1,t_1) = P_r \exp [- \int_{x_1,t_1}^{x_2,t_2} A_\mu dx^\mu]
\end{equation}
and taking the product of two such exponents, it is easy to see that
\begin{equation}
U_{r_1}(x_2,t_2; x_1,t_1) U_{r_2}(x_1,t_1; x_2,t_2)
= \exp [-{1 \over 2}\oint_{C} d\sigma^{\mu\nu} F_{\mu\nu}],
\end{equation}
using the Baker-Campbell-Hausdorff formula and the Stokes theorem, the
integration being done over the area enclosed by the closed path $r_1+r_2$.
As the curvature $F_{\mu\nu}$ vanishes,
\begin{equation}
U_{r_1}(x_2,t_2; x_1,t_1) U_{r_2}(x_1,t_1; x_2,t_2) = 1
\end{equation}
and so
\begin{equation}
U_r^{-1}(x_2,t_2; x_1,t_1) = U_r (x_1,t_1; x_2,t_2)
\end{equation}
so that
\begin{equation}
U_{r_1}(x_2,t_2; x_1,t_1) = U_{r_2}(x_2,t_2; x_1,t_1)
\end{equation}
ergo, $U$ is independent of the path taken. For a closed path,
$U(x,t; x,t)=1$. Hence path ordering drops out of the transition
matrix $T(x,y,\rho)$.
\par
        Returning to the time evolution of the transition matrix, it
can be shown that
\begin{equation}
\partial_t T(x,y,\rho) = [B(x,\rho), T(x,y,\rho)]
\end{equation}
which is the form of a Lax equation, so that all quantities of the form
\begin{equation}
K_n = {1 \over n} Tr [T(\rho)]^n; \quad K_0 = \ln [det T(\rho)]
\end{equation}
are conserved. We thus have an infinite number of conserved quantities
when the zero curvature conditions are fulfilled. That this holds also
for Toda Field Theories was shown by Olive and Turok [16].

\section{From Conformal Invariance To Toda Field Theory}
\setcounter{equation}{0}
\renewcommand{\theequation}{8.\arabic{equation}}

That the $KdV$ equation has a hidden conformal symmetry can be seen by
making a Fourier expansion with Fourier coefficients
\begin{equation}
u_n = - {1 \over 4} \int_{0}^{2\pi} u(x) {e^{-inx}} {dx \over {2\pi}}
\delta_{n0}
\end{equation}
It can be shown that the Poisson brackets of the $u_n$ satisfy the
Virasoro Algebra (up to trivial factors), i.e.,
\begin{equation}
-2i\pi \{u_n,u_m\} = - (n-m)u_{m+n} + {1 \over 2} n(n^2-1)\delta_{n+m}
\end{equation}
Higher order terms in the $KdV$ hierarchy have a hidden $\omega$ symmetry.
\par
        We now digress to take a look at Toda Field Theories. These are
essentially the only class of integrable, interacting, conformally invariant
field theories in two space-time dimensions. To see this, we start with
the generic action
\begin{equation}
S = \int [{1\over 2}\partial_\mu\phi_i\partial^\mu \phi_i-V(\phi_i)] d^2z
\end{equation}
The trace of the naive conserved energy-momentum tensor becomes
\begin{equation}
T_\mu^\mu = 2V.
\end{equation}
\par
        As the trace of the energy-momentum tensor is required to vanish
in a conformally invariant theory, it seems that if $V \neq 0$, the theory
is not conformal. However there is an ambiguity in the definition of the
energy-momentum tensor. If we attempt to improve the naive energy-momentum
tensor without violating the conservation property, we could choose
\begin{equation}
\theta_{\mu\nu} = T_{\mu\nu} + [\partial_\mu \partial_\nu - \eta_{\mu\nu}
{\partial}^2] f(\phi_i)
\end{equation}
whence the trace of the modified energy-momentum tensor is
\begin{equation}
\theta_{\mu\mu} = 2V + \partial_+ \partial_- f
\end{equation}
$\pm$ being the light cone directions. If the second term is to cancel the
first, we somehow need to get rid of the derivatives. This can be done,
using the equations of motion. Without knowing the explicit equations of
motion, the most general expression for $f(\phi_i)$ is $\sum c_i\phi_i$.
Using the equations of motion resulting from varying the action , the 
tracelessness condition becomes
\begin{equation}
2V + \sum_{i} c_i {{\partial V} \over {\partial \phi_i}} = 0
\end{equation}
Eq.(8,7) is easily solved, with the result that the trace of the
energy-momentum tensor vanishes if the potential is of the form
\begin{equation}
V(\phi_i) = \sum_{j} d_j {\exp [\sum b_{ij} \phi_{ji}]},
\end{equation}
satisfying the requirement
\begin{equation}
\sum_{i} c_i b_{ij} = -2
\end{equation}
We choose $b_{ij}$ to be related to the Cartan matrix of a simple
Lie Algebra. The resulting field theories are called Toda Field
Theories, and are described by the action
\begin{equation}
S_{Toda} = \int [{1\over 2}(\partial_\mu\phi,\partial^\mu\phi)
- {m^2 \over \beta^2} \sum {\exp {(\beta <\alpha^{(i)},\phi>)}}] d^2x
\end{equation}
where $< , >$ is the scalar product in the root space, and $\phi$
takes its values in the root space of the simple Lie Algebra on hand.
\par
        The equations of motion obtained from (8.10) are
\begin{equation}
\beta \partial^\mu \partial^\mu \phi_i + m^2 {\exp{(\sum K_{ij} \phi_j)}} =
0
\end{equation}
Specializing for the $SU(n)$ group, and setting $m=\beta=1$, this becomes
\begin{equation}
\partial_+ \partial_- \phi_i = - {\exp{(K_{ij} \phi_j)}}
\end{equation}
With $\phi_0 =0$ and $\phi_{i+1} =0$, this reduces to
\begin{equation}
\partial_+ \partial_- \phi_i = - {\exp{(2\phi_i-\phi_{i-1}-\phi_{i+1})}}
\end{equation}
Setting $$\psi_i = (\phi_i-\phi_{i-1})-(\phi_{i+1}-\phi_i),$$ after
Mikhailov [27], we get the equation
\begin{equation}
{\partial_t}^2 \psi_i - {\partial_x}^2 \psi_i = -[2{e^{\psi_i}}
- {e^{\psi_{i-1}}} - {e^{\psi_{i+1}}}]
\end{equation}
which is easily seen to be related to the Toda equations (5.20). One expects
that the Toda Field Theories are integrable, and it turns out that they
are indeed so (see ref.[8]). The calculation rests upon the existence of
a zero curvature condition for certain group theoretical combinations of
$\phi$, which can be chosen as gauge fields.
\par
        As mentioned earlier, the Toda Field Theories have been completely
solved for simple $g$ by Leonov and Saveliev [18]. They have also been
solved for affine $g$ by Olive and Turok [16].
\par
        Quantization of the Toda Field Theories is more problematic since
the potential has no local minimum, the latter being attained at infinity,
using the gauge group $A_1$. A lucid discussion of the problems encountered
in the theory is given in ref.[29].
\par
        The central charge of the Toda theories can be constructed using
free field technology, and is found to be [30]
\begin{equation}
C = {{\hbar r} \over {2\pi}} + 12 [{{\hbar\beta\rho} \over {4\pi}}
+ {{\rho^v} \over \beta}]^2,
\end{equation}
$r$ being the rank of the algebra, $\rho$ being half the sum of the
positive roots, and $\rho^v$ its dual. Eq.(8.15) gives an indication
that a quantum Toda theory with a strong coupling constant is equivalent
to another Toda theory with a weak coupling constant, obtained by replacing
$\beta$ by $4\pi/{\hbar\beta}$, and interchanging roots and "coroots".
\par
        Incidentally, strong/weak coupling duality has recently become a
subject of immense study in relation to string theories.
\par
        It is possible to obtain the minimal models from the Toda Field
Theories. For a particular value of $\beta$, the central charges can be
made to agree. However this is not enough. A complication arises from
the fact that not all primary fields in the minimal models are actually
present in the Toda theory. However, because of the duality in the theory,
we can add another part of the potential with the coupling constant
replaced by its dual; see Mansfield [31]. This modification is sufficient
to give complete agreement.

\section{W-Algebras: Hamiltonian Reduction of WZNW}
\setcounter{equation}{0}
\renewcommand{\theequation}{9.\arabic{equation}}

Another fact which makes the conformally invariant Toda theories
interesting is that to each such Toda theory, there corresponds a
$W$-algebra. The $W$-algebras are an extension of the Virasoro algebra
by adding primary fields primary fields of spin higher than $Z$, and were
introduced by Zamolodchikov [32] as a pointer to conformal field theories
with a larger overall symmetry. Zamolodchikov [32] investigated the case
in which a primary field $w(r)$ of weight $3$ is added to the Virasoro
algebra. In order for the algebra to be close, it had to be made
`non-linear', and hence lost its linear Lie Algebra character.
\par
        Balog et al [33-35] showed that the Liouville and Toda Field
Theories can be obtained as conformally reduced $WZNW$ theories. This
reduction can be viewed as a gauge procedure, and the Toda field theory
can be obtained as the gauge invariant content of a gauged $WZNW$ theory.
The Liouville theory is obtained for the special case of the $SL(2,R)$
gauge group.
\par
        The most powerful method of constructing $W$-algebras is through
the so-called quantum Drinfield-Sokolov reduction. In this, one starts
with an affine Lie Algebra, and reduces it by imposing some constraint on
its generators. At the classical level, this procedure which leads to the
so-called Gelfand-Dickey algebras [36], was pioneered by Drinfield and
Sokolov [37].
\par
        It is thus clear that under the reduction that takes a $WZNW$
field theory to a Toda field theory, the affine Lie Algebra that
characterizes the $WZNW$ theory reduces to a $W$-algebra that is associated
to a Toda field theory. This approach is also readily generalizable to the
supersymmetric case where various new $W$-superlagebras have been found as
symmetry algebras of supersymmetric Toda field theories. We refer the
interested reader to ref.[38] for further progress in this area.
\par
        In what follows, we review the essential steps of the Lagrangian
reduction of the $WZNW$ model. The $WZNW$ action for a non-compact group
$G$ in 2D Minkowski space-time is
\begin{equation}
S(g) = - {k \over {8\pi}} \int_{S^2} d^2\rho \eta^{\mu\nu}
Tr (g^{-1}\partial_\mu g)(g^{-1}\partial_\nu g)
+ {k \over {12\pi}} \int_{B} Tr (g^{-1} d g)^3
\end{equation}
where $B$ is the volume occupied by $S^2$. The left and right Affine
Kac-Moody [AKM] symmetries of this theory are generated by the
Noether currents
\begin{equation}
J(\lambda) = \kappa Tr[\lambda(\partial_+ g)g^{-1}]; \quad
{\tilde J}(\lambda) = -\kappa Tr[\lambda g^{-1}(\partial_- g)]
\end{equation}
where $\kappa$ = ${-k \over {4\pi}}$, and $\lambda$ is an element of
the Lie Algebra ${\bf g}$. The $WZNW$ equations of motion are known to
be equivalent to the current conservation
\begin{equation}
\partial_- J = \partial_+ {\tilde J} = 0.
\end{equation}
We now choose the following Gauss decomposition of an arbitrary element
$g$=$ABC$, e.g.,
\begin{eqnarray}
A &=& \exp [\sum_{\alpha \in \Delta^+} x^\alpha E_\alpha]; \\ \nonumber
B &=& \exp [{1\over 2}(\sum_{\alpha \in \Delta} \phi^\alpha H_\alpha)]; \\
\nonumber
C &=& \exp [\sum_{\alpha \in \Delta^-} y^\alpha E_\alpha];
\end{eqnarray}
where Cartan-Weyl root vectors $E_\alpha$, Cartan subalgebra generators
$H_\alpha$ = $[E_\alpha, E_{-\alpha}]$, and a set of positive (negative)
roots $\Delta^{\pm}$ have been introduced with the following properties
\begin{equation}
K_{\alpha\beta} =  \alpha(H_\beta) = {{2\alpha.\beta} \over
{\mid \alpha \mid}^2}; \quad {\alpha, \beta} \in \Delta; \quad
{\mid {\alpha_{long}} \mid}^2 =2 ;
\end{equation}
\begin{equation}
Tr(H_\alpha \dot H_\beta) = {2 \over {\mid \alpha \mid}^2} K_{\alpha\beta}
\equiv C_{\alpha\beta};
\end{equation}
\begin{equation}
Tr(E_\alpha \dot E_\beta) = {2 \over {\mid \alpha \mid}^2}
\delta_{\alpha, -\beta};  \quad Tr[E_\alpha, H_\beta]=0.
\end{equation}
We also introduce the Polyakov-Wiegmann identity
\begin{eqnarray}
S(ABC) &=& S(A)+S(B)+S(C)+ \kappa \int d^2\rho Tr [(A^{-1}\partial_-A)
\partial_+ B)B^{-1}   \\  \nonumber
       & & + (B^{-1}\partial_-B)(\partial_+C)C^{-1} + (A^{-1}\partial_-A)
(B(\partial_+C)C^{-1}B^{-1})]
\end{eqnarray}
We now see, using eqs.(9.4-9.8), that the generalized constraints
\begin{equation}
J(E_\alpha) = \kappa c_1^\alpha; \quad {\tilde J}(E_{-\alpha}
= -\kappa c_2^\alpha; \quad \alpha \in \Delta^+
\end{equation}
with some real numbers $c_{1,2}^\alpha$ whose values do not vanish only
for primitive roots $\alpha \in \Delta$, are enough to reduce the $G$-
based WZNW theory to the Toda Field Theory defined by the Lagrangian
\begin{equation}
L_{Toda} = -{k \over {8\pi}}[{1\over 4}C_{\alpha\beta} \partial_+\phi^\alpha
\partial_-\phi^\beta - \sum_{\alpha \in \Delta}(u^2)^\alpha
{e^{({1\over 2}K_{\alpha\beta} \phi^\beta)}}]
\end{equation}
where
$$ (u^2)^\alpha = {\mid \alpha \mid}^2 c_1^\alpha c_2^\alpha$$.
Due to $c_{1,2}^\alpha \neq 0$ for the primitive roots, the constraint
(9.7) can be re-written in terms of the Gauss decomposition (9.5-7)
as follows:
\begin{eqnarray}
A^{-1}\partial_- A &=& B[\sum_{\alpha \in \Delta}{1\over 2}{\mid \alpha
\mid}^2
c_2^\alpha E_\alpha]B^{-1}  \\  \nonumber
                   &=& \sum_{\alpha \in \Delta} {1\over 2} {\mid \alpha
\mid}^2
c_2^\alpha E_\alpha \exp [{1\over 2}K_{\alpha\beta} \phi^\beta];
\end{eqnarray}
\begin{eqnarray}
(\partial_+ C)C^{-1} &=& B^{-1}[\sum_{\alpha \in \Delta}{1\over 2}
{\mid \alpha \mid}^2 c_1^\alpha E_{-\alpha}]B  \\  \nonumber
                     &=& \sum_{\alpha \in \Delta} {1\over 2}
{\mid \alpha \mid}^2 c_1^\alpha E_{-\alpha}
\exp [{1\over 2}K_{\alpha\beta} \phi^\beta];
\end{eqnarray}
In the WZNW equations of motion, $A$ and $C$ occur only in the combinations
given in (9.11-12), so that they can be eliminated in favour of $B$ or
$\phi^\alpha$. The remaining equation is just the Toda equation [25,34,35]:
\begin{equation}
\partial_+\partial_-\phi^\alpha + {1\over 2}{\mid \alpha \mid}^2
(u^\alpha)^2
\exp [{1\over 2}K_{\alpha\beta} \phi^\beta] = 0;
\end{equation}
(see also ref.[25] for details).
\par
        As mentioned earlier, the Toda Field Theory possesses an extended
symmetry represented by a classical $W$-algebra. These $W$-algebras can be
obtained as the quantum versions of the so-called Gelfand-Dickey algebras
[36] known in the theory of $KdV$ equations. For instance, the Poisson
bracket associated with the $KdV$ equation in (8.2), results in the
classical version of the Virasoro algebra which is the simplest $W$-algebra.
Moreover, the Lax representation of the $KdV$ equation (3.15), defines the
third order differential operator $B=w^{(3)}$. The Fourier components of
$B$, along with those of the $KdV$ field, form the Gelfand-Dickey [36]
algebra that generalizes to $w^{(3)}$ in the quantum case.
\par
        Now regarding the Toda theory as a constrained $WZNW$ theory, the
Hamiltonian structure can be obtained by a classical Drinfield-Sokolov
reduction from the constrained phase space of the $AKM$ algebra. In the
Hamiltonian formalism, the $AKM$ symmetry of the $WZNW$ theory is
represented by first class constraints. The $W$-algebra of the Toda theory
arises as the Poisson bracket algebra of gauge-invariant polynomials of the
constrained $AKM$ currents and their derivatives. In what follows, we
summarize the arguments supporting these statements.
\par
        Let $g(z,{\bar z})$ be the $G$-valued $WZNW$ fields and $J(z)$ the
corresponding $AKM$ currents having the form
\begin{equation}
g(z,{\bar z})=g(z)g({\bar z}); \quad \partial g(z) = J(z)g(z)
\end{equation}
Let $dimg$ be the dimension of $G$; $l$ its rank; $k$ the level of the
associated $AKM$ algebra ${\hat g}$; $g$ the dual Coxeter number of $G$;
$\rho$ the half sum of the positive roots; and $\beta$ the dual of $\rho$.
\par
        The constrained $WZNW$ theory is specified by (9.9). After a
suitable choice of constants $c_i$, the currents $J(z)$ can be decomposed as
\begin{eqnarray}
J(z) &=& I_- + j(z); \quad I_- = \sum_{i=1}^{l} E_{-\alpha_i};  \\
\nonumber
j(z) &=& \sum_{i=1}^{l} j^{i}(z)H_i + \sum_{\phi \in \Delta^+} E_\phi
\end{eqnarray}
where $\{E_{\alpha_i}\}$ are $l$ simple roots of $g$. The maximal subgroup
of ${\hat G}$ leaving this form of currents invariant, is the maximal
nil-potent subgroup generated by $E\phi$, $(\phi \in \Delta^+)$, and
implemented by the $(dimg - l)/2$ constrained $AKM$ currents $J^\phi(z)$.
This allows us to interpret the constrained $WZNW$ theory as the gauge
theory
in which all but $l$ of the $(dimg + l)/2$ components of $J$ are
gauge components [33-35].
\par
        The current $j(z)$ and the gauge transformations corresponding to
$E_\phi$ act on each column of the $WZNW$ field $g(z)$ separately, while
each column contains only one gauge-invariant component $e$ (of the highest
weight), satisfying $E_\phi e = 0$. The gauge degrees of freedom
corresponding to the other elements of each column can be eliminated by a
gauge fixing in favour of $e$. Because of (9.15), this leads to a linear
pseudo-differential equation $De=0$, where $D$ is a polynomial
pseudo-differential operator whose coefficients are gauge invariant
polynomials in the currents $J$. This operator $D$ can now be used to
define a classical $W$-algebra by choosing a Drinfield-Sobolov gauge in
which one has
\begin{equation}
j_{DS} = \sum w^P (z) F_P
\end{equation}
where $P$'s are the orders of $l$ independent Casimir operators of $g$,
and $F_P$ generators with $H$ weights $(P-1)$, so that the gauge-fixed
current (9.16) has  only one non-vanishing component in each of the $l$
irreducible representations in a decomposition of the adjoint of $g$ w.r.t.
one of its sub-groups $SL(2,R)$. The Poisson brackets between the different
polynomials $w^P$ define a classical $W$-algebra.
\par
        We close this Section by noting that Toda field theories also play
an important role in the discussion of $W$-gravity, where they arise as
effective quantum theories [39,40] for the $W$-gravity degrees of freedom
in the conformal gauge. For a quantum version of the $WZNW \rightarrow Toda$
conformal reduction, see [34, 41].

\section{Self-Dual Y-M Theories: 2D Integrable Models}
\setcounter{equation}{0}
\renewcommand{\theequation}{10.\arabic{equation}}

The self-dual Yang-Mills (SDYM) theory appears to be a master theory for a
whole variety of 2D integrable systems, as we are now going to explain.
Though there is no general proof, the statement can be checked on a case
by case basis. The main point is that the 4D self-duality condition admits
of a zero curvature representation underlying a Hamiltonian description of
SDYM descendents in lower dimensions. This makes it possible to apply the
inverse scattering method for integration of the SDYM equations.
Simultaneously, it explains the origin of gauge symmetries in integrable
systems of the $KdV$ type, since the SDYM theory in both gauge and
conformally invariant in 4D. And last but not least, this connection
provides
us with a systematic way to associate the $KdV$ type hierarchy with any
simple Lie Algebra.
\par
        SDYM solutions invariant by the action of a subgroup with two
conformal generators satisfy a 2D differential equation, since eacd 1D
subgroup reduced the number of independent variables by one. This allows
us to describe the invariant SDYM solutions in terms of a 2D integrable
system. All known 2D integrable systems seem to be derivable this way, by
appropriate truncations of a 4D self-dual gauge theory. This is true, in
particular, for the $KdV$ and non-linear Schroedinger equations, the
Liouville and Toda equations, as well as other integrable in 2 and 3
dimensions. Our presentation in this Section is only illustrative; we
give one explicit example of embedding of the $KdV$ equation into the 4D
SDYM theory [42], and a supersummetric generalization.
\par
        Let $x^a$ = $(x,y,z,t)$ be the coordinates of a flat 4D space-time
of signature $(+,+,-,-)$. The invariant metric reads
\begin{equation}
ds^2 = 2 dx dz - 2 dy dt
\end{equation}
The SDYM equations in 2+2 dimensions $(\epsilon_{xyzt}=1)$ read as
\begin{equation}
F_{ab} = {1\over 2} \epsilon_{abcd} F^{cd}
\end{equation}
and are equivalently represented by 3 equations of the form
\begin{equation}
F_{tx} = F_{yz} = F_{ty} + F_{xz} = 0
\end{equation}
After a dimensional reduction which is equivalent to setting
\begin{equation}
\partial_y = \partial_z - \partial_x = 0,
\end{equation}
(10.3) takes the form
\begin{equation}
[\partial_t-H, \partial_x-Q]=[P,B]=0; \quad
[H,B] = [\partial_x -Q, \partial_x -P]
\end{equation}
where
$$ A_t =H; \quad A_x =Q; \quad A_y =-B; \quad A_z =P $$
It is clear that the first equation in (10.5) is a zero curvature equation.
We now choose the non-compact group and an embedding pattern in the form
\begin{equation}
B = \left(\matrix {
       0   &    0 \cr
       I   &    0
}\right) ;   \\
\end{equation}
\begin{equation}
Q = \left(\matrix {
       \lambda   &     1 \cr
        -u       &  - \lambda
}\right)
\end{equation}
where $\lambda$ is a constant and $u$ = $u(t, x+z)$. We can expand the
Lie Algebra-valued fields $H$ and $P$ as
\begin{equation}
H=H_-\tau_++H_+\tau_-+H_3\tau_3; \quad P=P_-\tau_++P_+\tau_-+P_3\tau_3
\end{equation}
where $\tau_{\pm}$ = $(\tau_1 \pm i\tau_2)/2$, and $\tau_{1,2,3}$ are
the Pauli spin matrices. It is clear that the second equation of (10.5)
gives
\begin{equation}
P_- = P_3 = 0,
\end{equation}
while the third equation of (10.5) gives
\begin{equation}
H_- =- P_+; \quad H_3 = -{1\over 2}{\partial_x (u+P_+^{'})} - \lambda P_+
\end{equation}
where primes denote derivatives w.r.t. $x$.
\par
        Finally, the first equation of (10.5) yields 3 equations
\begin{eqnarray}
H_+      &=& uP_+ - \lambda {\partial_x P_+} -{1\over 2}
{\partial_x\partial_x (u+P_+)}; \quad
{\partial_x (u +2P_+)} = 0;   \\  \nonumber
{\dot u} &=& {1\over 2}{\partial_x\partial_x\partial_x (u+P_+)}
+(u-P_+){\partial_x u} + 2\lambda^2 P_+
\end{eqnarray}
It follows that
\begin{eqnarray}
P_+  &=& -{1\over 2}u; \quad H_+ = -{1\over 2}u^2 + {\lambda \over 2}u_x
-{1\over 4} u_{xx};  \\  \nonumber
   {\dot u} &=& {1 \over 4} u_{xxx} +{3\over 2} u u_x - \lambda^2 u_x
\end{eqnarray}
Changing the notation as
$$ u \rightarrow u +{2\over 3}\lambda^2; \quad t \rightarrow 4t;  \quad
x+y \rightarrow x, $$
one obtains the $KdV$ equation
$$ u_t = u_{xxx} + 6u u_x. $$
This example may be relevant towards an ultimate unification of 2D
integrable models and 2D conformal field theories, as well as within
the 4D SDYM theories which are also closely related to $N+2$ strings.

\subsection{Self-Duality and Supersymmetry}

Extended Supersymmetry is compatible with self-duality in 2+2 dimensions.
Therefore the Supersymmetric self-dual Yang-Mills theory (SSDYM) is capable
of generating Supersymmetric 2D integrable models. However a Supersymmetric
generalization of the SDYM theory is not unique. One could either replace
a gauge group by its graded version, or a 2+2 dimensional space-time
by superspace.
\par
        Supersymmetric generalizations of the $KdV$ equation in 1+1
dimensions were obtained independently by Manin and Radul [43], Mathieu [44],
Bilal and Gervais [45]. These equations have two dynamical variables, one
bosonic $u(x,t)$, and one fermionic $\psi(x,t)$, and read
\begin{eqnarray}
{\partial_t u}   &=& {1\over 2} u_{xxx} + 3u {\partial_\lambda u}
+ {3\over 2} (\psi_{xx}) \psi    \\   \nonumber
{\partial_t \psi}&=& {1\over 2}\psi_{xxx} +{3\over 2}{\partial_x (u\psi)}
\end{eqnarray}
They are invariant under the $N=1$ Supersymmetry transformations
\begin{equation}
\delta u = \epsilon {\partial_x \psi};  \quad \delta \psi = \epsilon u
\end{equation}
$\epsilon$ being a constant Grassmann parameter. Eqs.(10.13) are integrable,
and can be obtained from the zero curvature condition associated with the
graded Lie Algebra $osp(2,1)$
\begin{equation}
{\partial_t A_x} - {\partial_x A_t} + [A_t, A_x] = 0
\end{equation}
when the following ansatz is used for 2D Yang-Mills potentials [45]:
\begin{equation}
2A_t(x,t) = \left(\matrix{
u_x     & u_{xx}+2u^2+\psi_x\psi & -i\psi_{xx}-2iu\psi\cr
-2u     &       - u_x                & i\psi_x\cr
i\psi_x &    i\psi_{xx} +2iu\psi            &   0
}\right)
\end{equation}
\begin{equation}
A_x = \left(\matrix{
  0    & u      & -i\psi\cr
-1    & 0      &    0\cr
  0    & i\psi  &    0
}\right)
\end{equation}
The 2D Super $KdV$ can be embedded into the self-duality equations by
choosing
the $osp(2/1)$-valued matrices $H,Q$ as $H=A_t(x,t)$, $Q=A_x(x,t)$, and
$B,P$
as $3\times 3$ matrices as
\begin{equation}
B = \left(\matrix{
  0    & {1\over 2}  & 0\cr
  0    &    0        & 0\cr
  0    &    0        & 0
}\right)
\end{equation}
\begin{equation}
P = \left(\matrix{
  0      & {u\over 2}       & -{3i \over 4}\psi\cr
  0      &     0            &         0\cr
  0      & {3i \over 4}\psi &         0
}\right)
\end{equation}
It can also be shown that the $N=1$ and $N=2$ Super $KdV$ equations,
as well as the $N=1$ Super-Liouville and Super-Toda equations, can all
be obtained from the $N=2$ SSDYM theory by dimensional reductions and
truncations [46]. A detailed analysis is however outside the scope of
this Article.

\section{Conclusions}

It has been our aim to present a bird's eye view of the important
developments in Integrable Systems over the past few decades. What
has been achieved is possibly a more subjective viewpoint, related to
building connections between sundry topics of immediate interest. It
has certainly not been possible to delve more deeply into the fascinating
developments in affine Toda Field Theory which seems to be a thrust area
of research today. We refer to the excellent lecture series by Corrigan [48]
on this subject. Neither is it possible to present an account of the
interesting link between the $KdV$ theory and Matrix models, encompassing
thus 2D gravity; (see ref.[25] for a readable account).Supersymmetric Toda
Field Theories have also been given the go by. They were first studied by
Evans and Hollowood [49], as well as by Leites et al [50]. It seems to be
possible to construct Toda Field Theories based on Lie Superalgebras with
one proviso, namely, that it is necessary that the Lie Superalgebra admits
a purely fermionic root system. This is only possible for the following
algebras:
$$ A(n,n-1);B(n,n);B(n-1,n);D(n.n-1);D(n,n); and D(2,1,\alpha).$$
In the generic case, $N=1$ Supersymmetric theories are obtained, which
can be formulated in $N=1$ superspace. There is one special case, namely,
the $sl(n,n-1)$ theories have in fact $N=2$ Supersymmetry; see [49].
Recently, Brink and Vasiliev [51] have proposed a model generalizing $A_N$
Toda Field Theories based on a continuous parameter, such that when this
parameter takes on certain discrete values, the model reduces to the
ordinary $A_N$ Toda Theories. More recently, Wyllard [52] has worked
out a $WZNW$ reduction of these generalized theories, and has also
attempted a Supersymmetric generalization [53] of the same.
\par
        One could also picture the affine Toda theories as integrable
deformations of the conformal Toda theories. As an example, by adding an
extra simple root to the $A_1$ Toda theory, one obtains the affine Toda
theory, which is also the $\sinh$-Gordon theory. General integrable
deformations have been investigated by Zamolodchikov, among others, as
an interesting field. Toda theories also appear in many other diverse
areas of theoretical physics, e.g., 1D discrete versions appear in the
physics of monopoles [54]. Further, certain 3D continuous Toda systems
are relevant to the classification of hyper-Kahler metrics in 4D [55].
Finally, it also appears that Toda Field Theories are relevant to
$M$-Theory, the conjectured all-in-all Theory encompassing all
String Theories [56].
\par
        I am deeply grateful to Prof. S.K. Malik, for providing me an
opportunity to write this topical Article on the subject of Toda Field
Theories. The literature on this subject is quite vast, but apart from
my own interest in Lie Algebras and super-algebras (which often draws me
into this area because of their obvious relevance to this subject), I
have benefitted greatly from some concentrated literature [22,25,56,57].
I am especially grateful to Prof. Ashok Das for his encouragement, and
for sparing time to go through the manuscript. I would also like to thank 
Jens Fjelstad for help with LaTeX.

\end{document}